\documentclass[a4paper,fleqn]{cas-sc}
\usepackage{placeins}
\usepackage{float}
\usepackage{makecell}

\usepackage[numbers]{natbib}
\def\tsc#1{\csdef{#1}{\textsc{\lowercase{#1}}\xspace}}
\tsc{WGM}
\tsc{QE}
\begin{document}
\let\WriteBookmarks\relax
\def\floatpagepagefraction{1}
\def\textpagefraction{.001}

\shorttitle{A new rotating machinery fault diagnosis method based on the Time Series Transformer}    

\shortauthors{Y. Jin, L. Hou and Y. Chen}

\title [mode = title]{A new rotating machinery fault diagnosis method based on the Time Series Transformer}  

\tnotemark[1] 

\tnotetext[1]{Supported by the National Natural Science Foundation of China (No. 11972129) and the National Major Science and Technology Projects of China (No. 2017-IV-0008-0045).}


\author[]{Yuhong Jin}[type=editor,
      style=chinese
]

% \cormark[1]


% \ead{<email address>}

% \ead[url]{<URL>}

% \credit{<Credit authorship details>}

% \affiliation[<aff no>]{organization={},
% %          citysep={}, % Uncomment if no comma needed between city and postcode

\author[]{Lei Hou}[type=editor,
      style=chinese,
      orcid=0000-0003-0271-7323
]

\cormark[1]

% Footnote of the second author

% Email id of the second author
\ead{houlei@hit.edu.cn}

\ead[url]{http://homepage.hit.edu.cn/houlei}

\author[]{Yushu Chen}[type=editor,
      style=chinese
]

% \credit{1111}

% \affiliation[1]{organization={11},
%             addressline={11}, 
%             city={11},
% %          citysep={}, % Uncomment if no comma needed between city and postcode
%             postcode={111}, 
%             state={11},
%             country={11}}

\cortext[1]{Corresponding author}

% \fntext[1]{111}

\affiliation[]{organization={School of Astronautics},
            addressline={Harbin Institute of Technology}, 
            city={Harbin},
%          citysep={}, % Uncomment if no comma needed between city and postcode
            postcode={150001}, 
            % state={11},
            country={P. R. China}}


\begin{abstract}
      Fault diagnosis of rotating machinery is an important engineering problem. In recent years, fault diagnosis methods based on the Convolutional Neural Network (CNN) and Recurrent Neural Network (RNN) have been mature, but Transformer has not been widely used in the field of fault diagnosis. To address these deficiencies, a new method based on the Time Series Transformer (TST) is proposed to recognize the fault mode of bearings. In this paper, our contributions include: Firstly, we designed a tokens sequences generation method which can handle data in 1D format, namely time series tokenizer. Then, the TST combining time series tokenizer and Transformer was introduced. Furthermore, the test results on the given dataset show that the proposed method has better fault identification capability than the traditional CNN and RNN models. Secondly, through the experiments, the effect of structural hyperparameters such as subsequence length and embedding dimension on fault diagnosis performance, computational complexity and parameters number of the TST is analyzed in detail. The influence laws of some hyperparameters are obtained. Finally, via t-Distributed Stochastic Neighbor Embedding (t-SNE) dimensionality reduction method, the feature vectors in the embedding space are visualized. On this basis, the working pattern of TST has been explained to a certain extent. Moreover, by analyzing the distribution form of the feature vectors, we find that compared with the traditional CNN and RNN models, the feature vectors extracted by the method in this paper show the best intra-class compactness and inter-class separability. These results further demonstrate the effectiveness of the proposed method.
\end{abstract}


\begin{highlights}
\item A generation method of tokens sequence which can process 1D format data is introduced.
\item A new fault diagnosis model based on the time series tokenizer and Transformer is proposed. 
\item The effect of structural hyperparameters on fault diagnosis performance, computational complexity and parameters number of the TST is discussed in detail.
\item t-Distributed Stochastic Neighbor embedding (t-SNE) is adopted to visualize the feature vectors distribution in the embedding space. 
\end{highlights}

% Keywords
\begin{keywords}
      Rotating machinery \sep Bearing fault diagnosis \sep Transformer \sep Deep learning
\end{keywords}

\maketitle

% Main text
\section{Introduction}
Rotating machinery has a pivotal role in modern industry, which is widely used in many industrial fields including aviation, aerospace engineering, naval architecture, motor industry and so on \cite{ZhaoXL}. A good deal of mechanical equipment such as aero-engine, gas turbines and wind turbines are inseparable from rotating machinery. However, due to the harsh operating environment, rotating machinery is prone to failure, and according to statistics, about 45\%-55\% of rotating machinery and equipment failure is caused by the damage to the bearing part \cite{Nandi}. Therefore, accurate bearing fault diagnosis is of great significance to improve the reliability and operation safety of rotating machinery \cite{Hoang}. In view the bearing fault diagnosis problem, many researchers have conducted a lot of research from multiple aspects. Among them, with the development of various intelligent algorithms in Machine Learning (ML), data-driven fault diagnosis method has been welcomed by more and more researchers \cite{Gao}. Currently, ML methods commonly utilized in fault diagnosis field include Support Vector Machine (SVM), K-Nearest Neighbor (KNN) algorithm, Self-Organized Map (SOM), etc. Yan and Jia \cite{YAN201847} adopted Variational Mode Decomposition (VMD) method to process the vibration signals of bearings, and the high-dimensional time-frequency domain features are extracted by combining the time-domain and frequency-domain features. Laplacian Score (LS) is employed to reduce the high dimension of features. Then, the features after dimension reduction are used as the input. Based on SVM and Particle Swarm Optimization (PSO) optimizer, a bearing fault diagnosis method is proposed. Dong et al. \cite{Dong} presented a bearing fault diagnosis method combining the Fuzzy C-means Method (FCM) and KNN, in which KNN also adopts the PSO algorithm for optimization to reduce the computation complexity. The experimental results show that this method can accurately identify the main bearing fault types with fewer data. Based on SOM network and Kurtosis criterion VMD method, Xiao et al. \cite{Xiao} realized a more accurate fault diagnosis result for an actual gear system. Such approaches, however, features are artificially extracted from the original vibration signals using various signal processing algorithms, which do not take full advantage of  the intelligent algorithms in ML. 

In recent years, Deep Learning (DL), as a very important branch of ML, has developed rapidly, and has made great achievements in many fields such as Computer Vision (CV), Natural Language Processing (NLP) and Autonomous Driving (AD) \cite{ZHAO2019213}. DL algorithms represented by various neural networks have powerful feature extraction capabilities, which can perform automatic representation learning from a lot of data and have strong adaptation \cite{DLSurvey}. DL models such as Stacked Auto-Encoder (SAE), Deep Belief Network (DBN), Convolutional Neural Network (CNN) and Recurrent Neural Network (RNN) have also gained many applications in the rotating machinery fault diagnosis field. Shao et al. \cite{ShaoSAE} presented a novel SAE feature learning method based on generalized correlation entropy and Hessian sparse regularization \cite{LIU201659}, which is robust to signal complexity and background noise. Gan et al. \cite{GAN201692} established a deep feature extraction approach suitable for various mechanical systems by stacking two-layer DBN models in a specific order, which is denoted as Hierarchical Deep Network (HDN). This method achieves better fault diagnosis results than Back Propagation Neural Network (BPNN) and SVM. Janessens et al. \cite{JANSSENS2016331} introduced the CNN model to recognize the health conditions of rotating machinery with four states. The input of network is the frequency-domain signals obtained by the Discrete Fourier Transform (DFT), and the CNN model consists of a convolution layer and a fully connected layer. Wang et al. \cite{WANG2017310} proposed a gearbox fault diagnosis method based on the CNN model. Raw vibration signals from sensors are converted into $60 \times 60$ 2D time-frequency images by the wavelet analysis and interpolation method, and then it is input into a CNN with two convolution layers. The experimental results show that the recognition accuracy of this method can achieve 99.58\% under the given dataset. Meanwhile, the influence of the number of filters on CNN performance is also discussed. The analysis results illustrate that under the network architecture and dataset in reference \cite{WANG2017310}, CNN has the best performance when the number of filters is set to 6. Lu et al. \cite{LU2017139} reformatted the original time series data with 400 length into a matrix of $20 \times 20$ utilizing means of sliding insertion, so that it can be processed by the 2D format CNN model. In literature \cite{7880628}, Ding et al. combined basic CNN model with the multiple scale layer to diagnosis fault mode of rolling bearings, which employs Wavelet Packet Energy (WPE) as input. Compared with the traditional CNN, this method presents better feature extraction ability. Guo et al. \cite{GUO2016490} stacked LeCun's LeNet5 model \cite{726791}, and proposed a two-layer hierarchical CNN model, whose input is the $32 \times 32$ matrix after reforming the original time series data of length 1024. The first layer of the model is employed to identify the fault type, and the second layer is used to further calculate the defect size of the fault. Since the vanilla CNN model is utilized to process images, many researchers will first convert the original vibration signal into 2D format data when using CNN for fault diagnosis. Nevertheless, it should be noted that CNN is also able to dispose 1D format data directly. In literature \cite{7501527}, Ince et al. made use of the 1D format CNN model to make fault classification based on the real-time data, which still achieves great fault diagnosis results. In the field of CV, small convolution kernels are usually adopted when designing CNN. However, Zhang et al. pointed out that when using 1D format CNN for fault diagnosis, if all convolutional layers use small convolution kernels, the network will have poor performance and can be easily affected by environmental noise. In terms of this issue, Zhang et al. \cite{ZHANG2018439} proposed Deep Convolutional Neural Networks with Wide First-layer Kernels (WDCNN) model. The first convolution layer of this model has a wide convolution kernel with a size of 256, which is used to suppress the influence of background noise, and then several small convolution kernels are employed for feature extraction and expression. The experimental results show that the WDCNN model is robust in noisy environment. Besides CNN, RNN can effectively model the sequence data \cite{6796344}, which is also widely utilized in the fault diagnosis field. The vanilla RNN model has problems of long-range dependent defects and gradient disappearance. To solve these problems, researchers proposed the Gated Recurrent Unit (GRU), Long short-term Memory Unit (LSTM) and other improved RNN models. Yuan et al. \cite{7748035} studied the problem of aero-engine fault diagnosis and residual life prediction on NASA's fault dataset based on the vanilla RNN, GRU and LSTM. Zhao et al. \cite{ZhaoRui2} designed a more complex deep learning model by combining CNN and LSTM, denoted as Bi-directional Long Short-Term Memory Networks (CBLSTM). Firstly, a convolutional layer is adopted to extract local features from the original vibration signals, and then the local features are input into a multi-layer bidirectional LSTM network to learn global features. With this stacked approach, CBLSTM performs better than many baseline approaches. Previous studies indicate that the fault diagnosis method based on CNN and RNN has been quite mature. However, there are still some shortcomings in CNN and RNN. The Pooling layer in CNN will lead to the defects of translation invariance (not equivalence) and information loss. In addition, the pooling operation can cause the model to ignore the association between the whole and the part. But removing the pooling layer will bring about a new problem that the receptive field of neuron is too small. In terms of RNN, neither vanilla RNN LSTM nor GRU has completely solved the problem of long-term dependencies, indicating RNN still has disadvantages in the modeling of long sequences. Then, RNN is also difficult to parallelize due to its inherent characteristics of calculation state and therefore hard to expand. 

Recently, to figure out the problems in the traditional DL models, Attention Mechanism (AM) \cite{AM1, bahdanau2016neural} is introduced. AM is a technology that can model sequence dependencies, which can be adopted in various DL models. The neural network architectures combining AM with CNN and RNN have also been established, such as Squeeze-and-Excitation Network \cite{8701503} and Structured Attention Networks (SAN) \cite{kim2017structured}. Furthermore, in 2017, Google Brain's Vaswani et al. came up with a new network architecture called Transformer \cite{Transformer}, which has made AM more widely available. Transformer architecture completely abandons CNN and RNN, and only contains Multi-head Self-Attention (MSA) mechanism and basic fully connected layer. This model attained the best results on the machine translation task at that time. Then, based on Transformer, Jacob et al. \cite{BERT} proposed the BERT algorithm to generate word vectors, which achieved a significant improvement in 11 various NLP tasks, including text classification, machine translation, and semantic segmentation. At present, Transformer has almost completely replaced RNN model in NLP field, and meanwhile, it is gradually gaining applications in CV. In 2021, Dosovitskiy et al. \cite{ViT} offered a model that applies Transformer to image classification tasks, namely Vision Transformer (ViT). Similar to the word embedding process in NLP, ViT divides the images into several fixed size patches, and then patch embedding is carried out by using linear transformation. Finally, the patch embedding results can be directly input into Transformer for feature extraction and classification. According to the test results, since CNN has inductive bias such as translation equivalence in image processing, the performance of ViT will be slightly worse than ResNet \cite{ResNet} if it is trained from scratch on the ImageNet dataset. However, if ViT is fine-tuned after pre-training on a larger dataset (JFT-300M), the performance of it will be able to exceed state-of-the-art method (ResNet152x4). Besides, the experimental results also show that ViT has strong scalability. When the data amount and model scale increase, its performance will continue to improve. Moreover, ViT has better parallelism in computing, which has greater advantages in processing large-scale data. Then, based on ViT, DL researchers also proposed a variety of modified visual Transformer models, such as CaiT \cite{CaiT}, T2TViT \cite{T2TViT}, DeiT \cite{DeiT}, CrossViT \cite{CrossViT} and PVT \cite{PVT}. However, although Transformer has made a lot of progress in CV and NLP, it has not been widely adopted in the field of fault diagnosis. 

According to the current research situation, the fault diagnosis method based on CNN and RNN has been well studied, but there are some defects of the CNN and RNN models. Nevertheless, the Transformer architecture that addresses these shortcomings has not gained widely applications in the field of fault diagnosis. In view of these deficiencies, to better solve the rotating machinery fault diagnosis problem, in this study, a new bearing fault diagnosis model based on the Time Series Transformer (TST) is proposed. Considering that most of the actual vibration signals are time series data in 1D format, we designed a new generation mode of tokens sequences called time series tokenizer. In addition, Transformer is introduced as the feature extraction module. On this basis, combining the Transformer and time series tokenizer, a new fault diagnosis model without CNN and RNN architectures is established, which can directly process 1D format vibration signals. Experimental results show that TST has preferable fault diagnosis performance compared with the traditional ML and DL approaches. Besides, the influence of structural hyperparameters on the recognition accuracy, computational complexity and the number of learnable parameters of the proposed TST is analyzed in detail. Finally, the feature vector is visualized and the distribution form of it is discussed. The analysis results shows that the feature vectors extracted by TST have better intra-class compactness and inter-class separability, which further proved the effectiveness of the proposed method in this paper. 

\section{Time Series Transformer}
\begin{figure}[h]
	\centering
		\includegraphics[width = 0.9\textwidth]{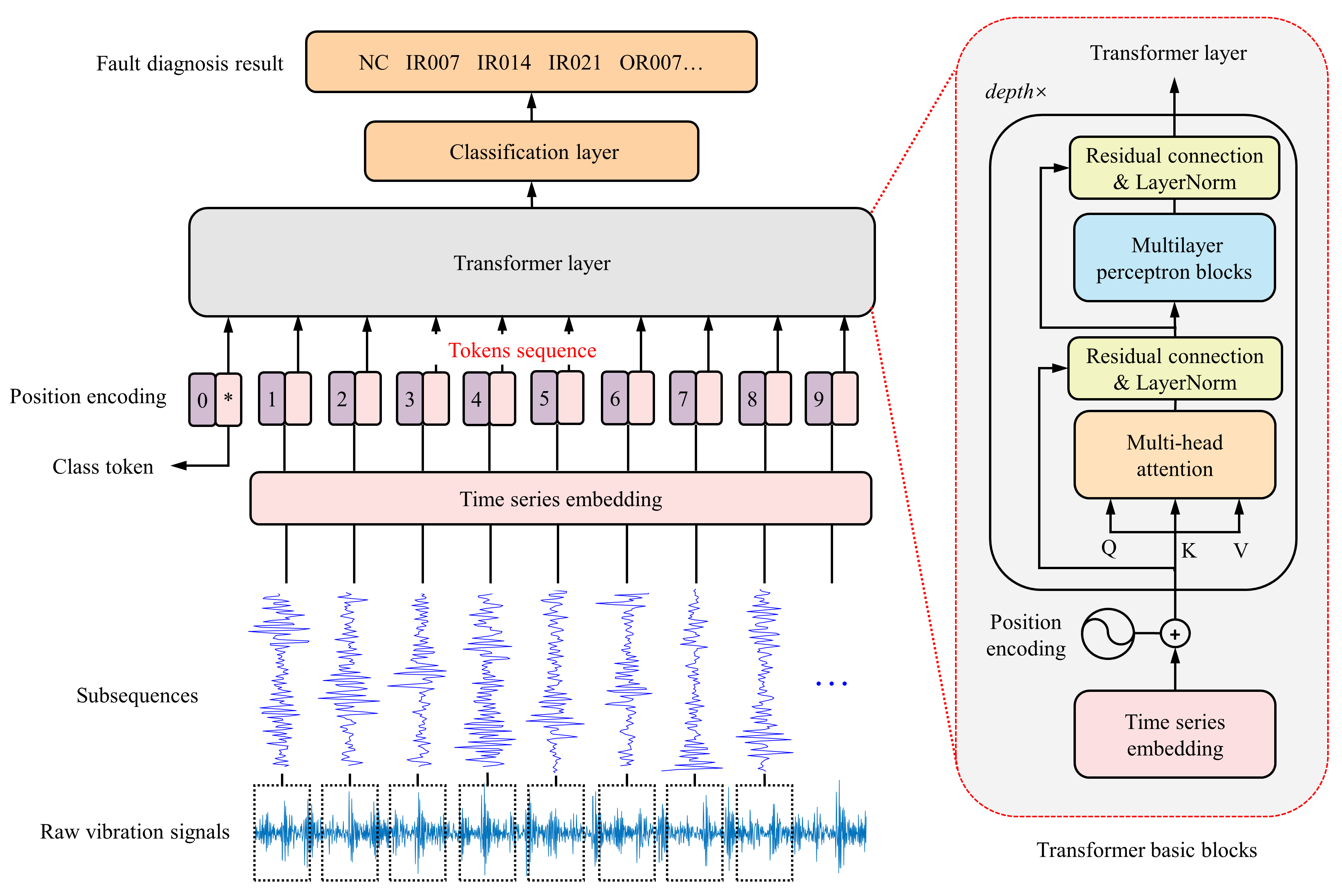}
	  \caption{Structure of the proposed TST model}
        \label{fig:TSTModel}
\end{figure}

The data format of raw vibration signals is generally a time series. However, the input of the vanilla Transformer model used in NLP is a text sequence after word segmentation, and the process object of the visual Transformer (ViT) adopted in CV is generally a 3-channel RGB image, so vibration signals cannot be processed directly. In this paper, to carry out rotor system fault diagnosis based on the Transformer, we propose the Time Series Transformer (TST) that can directly process data in 1D format. As shown in Fig.\ref{fig:TSTModel}, the proposed TST model is characterized by time series tokenizer, Transformer layer and classification layer. This section will discuss the details of these component.

\subsection{Time series tokenizer}
Referring to the concept in NLP, the input of Transformer is called tokens sequence. The proposed TST utilizes the time series tokenizer to obtain the tokens sequence from the raw vibration signals. 

\subsubsection{Time series embedding}
In principle, time series data with batches can be denoted as $t \in \mathbb{R}^{B \times L}$, where $B$ is batch size and $L$ is the length of the given vibration signal time series. During the process of time series embedding, firstly, similar to word segmentation in NLP, the input time series is trimmed into several subsequences with the given length and pieced together into a tensor denoted as $[t_{s}^{1}, t_{s}^{2}, t_{s}^{3},..., t_{s}^{N_{s}}] \in \mathbb{R}^{B \times N_{s} \times (L/N_{s})}$, where $N_{s}$ is the number of divided subsequences. Both $N_{s}$ and $L/N_{s}$ must be integer. Then, like word embedding, the subsequences is mapped to the higher-dimensional embedding space by a simple linear transformation, which can be given as Eq.\ref{eq:TimeSeriesEmbedding}
\begin{equation}
      TokensSeq = [t_{s}^{1}, t_{s}^{2}, t_{s}^{3},..., t_{s}^{N_{s}}] \cdot W_{embedding} \in \mathbb{R}^{B \times N_{s} \times dim}
      \label{eq:TimeSeriesEmbedding}
\end{equation}
where $W_{embedding} \in \mathbb{R}^{(L/N_{s}) \times dim}$ is a learnable matrix and $dim$ is the dimension of the time series embedding. It should be pointed that in order for TST to learn more general mapping, all subsequences share the same linear mapping matrix. Furthermore, the whole process of time series embedding is equivalent to a multi-channel 1D convolution operation on the input time series, but it does not mean that the proposed TST contains the convolution layer.

In fact, time series embedding can also be regarded as the process of extracting features from the input vibration signal. Similar to the method of embedding the input image with different patch size in CrossViT\cite{CrossViT}, when the length of the subsequences is small, time series embedding can better extract the local features of the time series; conversely, time series embedding may concentrate on the global features.

\subsubsection{Class token}
The proposed TST needs to represent the features after extracting them from the tokens sequence. There are two main approaches: one is to carry out global pooling of the last Transformer layer; the other is to obtain the feature map by adding class token in the tokens sequence referring to BERT\cite{BERT}. The class token a randomly initialized learnable sequence denoted as $x_{0} \in \mathbb{R}^{1 \times dim}$, and the tokens sequence with class token is defined as
\begin{equation}
      TokensSeq = [x_0; [t_{s}^{1}, t_{s}^{2}, t_{s}^{3},..., t_{s}^{N_{s}}] \cdot W_{embedding}] \in \mathbb{R}^{B \times (N_{s}+1) \times dim}
      \label{eq:ClassToken}
\end{equation}
Eq.\eqref{eq:ClassToken} show that after the computation of the multi-head attention mechanism, $x_{0}$ is related to all subsequences, which means that $x_{0}$ fuse the features from all subsequences, so $x_{0}$ can be adopted as the feature map.

\subsubsection{Position encoding}
Since there is no convolution operation in the proposed TST, and the multi-head self-attention mechanism does not contain the position information within the tokens sequence during calculating, position encoding is added to retain the absolute and relative position information in the tokens subsequence, which can be obtained by Eq.\ref{eq:PositionEncoding}
\begin{equation}
      TokensSeq = [x_0; [t_{s}^{1}, t_{s}^{2}, t_{s}^{3},..., t_{s}^{N_{s}}] \cdot W_{embedding}] + E_{pos} \in \mathbb{R}^{B \times (N_{s}+1) \times dim}
      \label{eq:PositionEncoding}
\end{equation}
where $E_{pos} \in \mathbb{R}^{(N_{s}+1) \times dim}$. Reference \cite{ViT} indicates that there are two formats for position encoding: 1D and 2D. In 1D position encoding, the tokens sequence is considered as a sequence of subsequences arranged in order. $E_{pos} \in \mathbb{R}^{(N_{s}+1) \times dim}$ is a randomly initialized learnable matrix and the number of learnable parameters is ${(N_{s}+1) \times dim}$. This position encoding format is more suitable for the time series. Besides, 2D position encoding regards the inputs as a grid of patches in two dimensions, which means that $E_{pos} \in \mathbb{R}^{(N_{s}+1) \times (dim/2) \times (dim/2)}$. But the results in reference \cite{ViT} show that they have not observed significant performance gains from using 2D position encoding and this method is not applied to the 1D format time series. Of course, the tokens sequence can also just be considered as a bag of subsequences which indicate that there is no position encoding. The influence of position encoding will be analyzed in Section 4. 

\subsection{Transformer layer}
The Transformer layer is the core part of the proposed TST for feature extraction and representation, which consists of $N$ Transformer basic blocks stacked on top of each other. The output of the $l$th Transformer basic block is adopted as the input of the next block. The basic form of the Transformer layer in the TST is similar to the vanilla Transformer in reference \cite{Transformer}, mainly including multi-head self-attention mechanism (MSA), multilayer perceptron blocks (MLP). Besides, the proposed TST also make some improvements referring to ViT and the details are as follows.

\subsubsection{Multi-head self-attention mechanism}
Multi-head self-attention mechanism is the most critical definition in Transformer basic blocks, which is based on the attention mechanism. An attention mechanism can be regarded as a mapping from a query ($Q$) and a set of key ($K$)-value ($V$) to an output, where $Q$, $K$, $V$, and output are all vectors. The output is actually a weighted sum over $V$, and the weight matrix (called Attention Distribution, AD) is calculated from $Q$ and $K$. The essence of AD is to characterize the sequence similarity and there are many methods (denoted as score function) to obtain it. Commonly used score function include Scaled Dot-Product Attention, Bahdanau Attention \cite{bahdanau2016neural}, Content-base Attention, etc. Among them, Scaled Dot-Product Attention is widely utilized by researchers because it is simple to calculate, easy to parallelize and does not introduce additional parameters in the model. Scaled Dot-Product Attention is described as follows
\begin{equation}
      Attention(Q, K, V) = softmax(\frac{QK^{T}}{\sqrt{d_{k}}})V
      \label{eq:SDPAttn}
\end{equation}
where $d_{k}$ is the dimension of queries and keys. $\frac{1}{\sqrt{d_{k}}}$ in Eq.\eqref{eq:SDPAttn} is the scaling factor which plays a role of stabilizing the gradient. Attention mechanism can make each token in the tokens sequence tend to get information
from others, which is useful for dealing with the long sequences problem in NLP. However, classical attention mechanism like Eq.\eqref{eq:SDPAttn} cannot obtain the information from different subspaces. To solve this deficiency, researchers have proposed multi-head attention mechanism. Similar to partial convolution operation in AlexNet, multi-head attention mechanism can be given as
\begin{equation}
      \begin{aligned}
            MultiHead(Q, K, V) &= Concat(Head_{1}, Head_{2},..., Head_{h})W_{O}\\
            \text{where} Head_{i} &= Attention(QW_{Q}^{i}, KW_{K}^{i}, VW_{V}^{i})
      \end{aligned}
      \label{eq:MHA}
\end{equation}
where $W_{O} \in \mathbb{R}^{hd_{v} \times dim}$, $W_{Q}^{i} \in \mathbb{R}^{dim \times d_{k}}$, $W_{K}^{i} \in \mathbb{R}^{dim \times d_{k}}$, $W_{V}^{i} \in \mathbb{R}^{dim \times d_{v}}$, $h$ is the number of $Head$ and $d_{v}$ is the dimension of values. 

In NLP, $Q, K, V$ in Eq.\eqref{eq:MHA} are generally different. However, when the $Q, K, V$ are all the same, the multi-head attention mechanism can be used as a feature extraction algorithm, namely MSA. The output of MSA in $l$th Transformer block with the layer normalization (LayerNorm) and residual connection, denoted as $y_{l}^{MSA} \in \mathbb{R}^{B \times (N_{s}+1) \times dim}$, can be defined as follows
\begin{equation}
      \begin{aligned}
            y_{l}^{MSA} &= MSA(LayerNorm(y_{l-1})) + y_{l-1}\\ 
            &= MultiHead(LayerNorm(y_{l-1}), LayerNorm(y_{l-1}), LayerNorm(y_{l-1})) + y_{l-1}
      \end{aligned}
      \label{eq:MSAOut}
\end{equation}
where $y_{l}$ is the output of the $l$th Transformer basic block and $y_{0} = TokensSeq$ in Eq.\eqref{eq:PositionEncoding}. Besides, $l = 1, 2, 3,..., depth$ and $depth$ indicates the number of stacked Transformer basic blocks in the Transformer layer, which is an adjustable hyperparameter in the proposed TST. 

\subsubsection{Multilayer perceptron blocks}
To make the proposed TST achieve more complex nonlinear mapping and introduce the local and translational equivariant, MLP is also included in the Transformer layer. MLP in $l$th Transformer layer consists of one nonlinear transformation with an activation function and one linear transformation which is described in Eq.\eqref{eq:MLP}
\begin{equation}
      MLP(y_{l}^{MSA}) = activation(y_{l}^{MSA}W_{1}^{l}+b_{1}^{l})W_{2}^{l}+b_{2}^{l}
      \label{eq:MLP}
\end{equation}
where $W_{1}^{l} \in \mathbb{R}^{dim \times dim_{MLP}}$, $b_{1}^{l} \in \mathbb{R}^{dim_{MLP}}$, $W_{2}^{l} \in \mathbb{R}^{dim_{MLP} \times dim}$, $b_{1}^{l} \in \mathbb{R}^{dim}$. $dim_{MLP}$ represents the embedding dimension of the nonlinear transformation in MLP so $dim_{MLP}$ is greater than $dim$, generally. From Eq.\eqref{eq:MLP} we know that the MLP first adopt the nonlinear layer to perform nonlinear dimension raising operation on the input, and then utilize the linear layer to perform a linear dimension reduction operation, which is a classic feature extraction method. Besides, to improve the performance of the TST, the activation function used in the MLP is Gaussian error Linear Unit (GeLU) rather than Rectified Linear Unit (ReLU), as shown in Eq.\eqref{eq:GeLU}
\begin{equation}
      GeLU(x) = xP(X \leq x) = x \phi (x) = x[1 + erf(x/{\sqrt{2}})]/2 \approx 0.5x(1 + tanh[\sqrt{2/\pi}(x + 0.045x^3)])
      \label{eq:GeLU}
\end{equation}
where $\phi (x)$ is the standard Gaussian distribution function. Instead of the sign function, GeLU employs the Gaussian distribution of the input to calculate weights. GeLU can be regarded as a smooth version of ReLU avoiding the shortage of derivative discontinuity at 0. Moreover, GeLU also has gradient in the negative part, solving the problem of 'dying ReLU'. 

In the proposed TST, MLP also combined with the LayerNorm and residual connection. The output of MLP in the $l$th Transformer block (namely $y_{l}^{MLP}$) can be obtained by Eq.\eqref{eq:MLP2}
\begin{equation}
      y_{l}^{MLP} = MLP(LayerNorm(y_{l}^{MSA})) + y_{l}^{MSA}
      \label{eq:MLP2}
\end{equation}
Based on Eq.\eqref{eq:MSAOut} and Eq.\eqref{eq:MLP2} the output of the entire Transformer layer $y$ is described by
\begin{equation}
      \begin{aligned}
            y_{l}^{MSA} &= MSA(LayerNorm(y_{l-1})) + y_{l-1},        & l = 1, 2, 3,..., depth\\
            y_{l}^{MLP} &= MLP(LayerNorm(y_{l}^{MSA})) + y_{l}^{MSA},        & l = 1, 2, 3,..., depth\\
            y &= LayerNorm(y_{depth}^{MLP}[0])
      \end{aligned}
      \label{eq:TransformerLayerOut}
\end{equation}
where $y_{depth}^{MLP}[0]$ denotes the class token (shown in Eq.\eqref{eq:ClassToken}) of the last Transformer basic block, which is the feature map extracting from the input vibration signal.

\subsection{Classification layer}
Classification layer is introduced to transform the feature map extracted by the Transformer layer into one-hot encoding for pattern recognition. Its basic structure is similar to MLP in Eq.\eqref{eq:MLP}, except for the addition of the Softmax function.
\begin{equation}
      ClassLayer(y) = y^{class} = Softmax(yW^{class}+b^{class})
      \label{eq:ClassLayer}
\end{equation}
where $W^{class} \in \mathbb{R}^{dim \times N_{class}}$, $b^{class} \in \mathbb{R}^{N_{class}}$, $N_{class}$ is the number of failure mode. We can obtain the probability value that the input vibration signal belongs to each category by the classification layer, so that the failure mode of samples can be gained.

\subsection{Other details in the proposed TST}
The loss function adopted in the TST is the cross-entropy (denoted as $L_{ce}$), which is commonly utilized for the pattern recognition problem. The formula of $L_{ce}$ is defined as Eq.\eqref{eq:CE}
\begin{equation}
      {L_{ce}(\theta)} =  - \frac{1}{B}\left[ {\sum\limits_{i = 1}^B {\sum\limits_{j = 1}^{{N_{class}}} \chi  } \left\{ {category({{t}^{(i)}}) = j} \right\}\log (y^{class}[j])} \right]
      \label{eq:CE}
\end{equation}
where $\chi$ is the indicative function, ${t}^{(i)}$ is the $i$th sample and $y^{class}[j]$ denotes the probability value that ${t}^{(i)}$ belongs to $j$th category. $\theta$ is the learnable parameters in the TST and Adam optimizer is used to minimize $L_{ce}(\theta)$. Besides, to perform more refined searches, the decaying learning rate schedule is carried out in the optimizer where step size is set to 10 and the damping factor $\gamma$=0.8. Moreover, Dropout is introduced in each Transformer basic block and position encoding process aiming at avoiding overfitting during training. By discarding the current parameters with a certain probability (namely $p_{drop}$), Dropout adds a strong regularization constraint to the TST. 

Finally, the concrete parameters of the proposed TST are shown in Tab.\ref{tbl:TSTParameters}. The adjustable hyperparameters in the proposed TST include $N_{s}$, $dim$, the position encoding format, $d_{k}$, $h$, $dim_{MLP}$, $depth$ and $p_{drop}$. The effects of some hyperparameters will be discussed in detail later in this paper. 
\begin{table}[H]
      \caption{Concrete parameters of the proposed TST}\label{tbl:TSTParameters}
      \begin{tabular*}{1.0\textwidth}{@{}LL@{}}
            \toprule
            Parameters & Value \\ % Table header row
            \midrule
            Input time series length $L$ & 2048$\times$1 \\
            Subsequences length $L/(N_{s})$ & 8 \\
            The dimension of the time series embedding $dim$ & 128 \\
            Position encoding format & 1D \\
            The number of stacked Transformer basic blocks in the Transformer layer $depth$ & 6 \\
            The number of $Head$ in MSA $h$ & 6 \\
            The dimension of queries and keys in MSA $d_{k}$ & 64 \\
            The embedding dimension of the nonlinear transformation in MLP $dim_{MLP}$ & 256 \\
            Dropout probability $p_{drop}$ & 0.1 \\
            Initial learning rate of the optimizer &  3$\times$\text{$10^{-5}$} \\
            \bottomrule
      \end{tabular*}
\end{table}

\section{Brief introduction to dataset}
\begin{figure}[h]
	\centering
		\includegraphics[width = 0.8\textwidth]{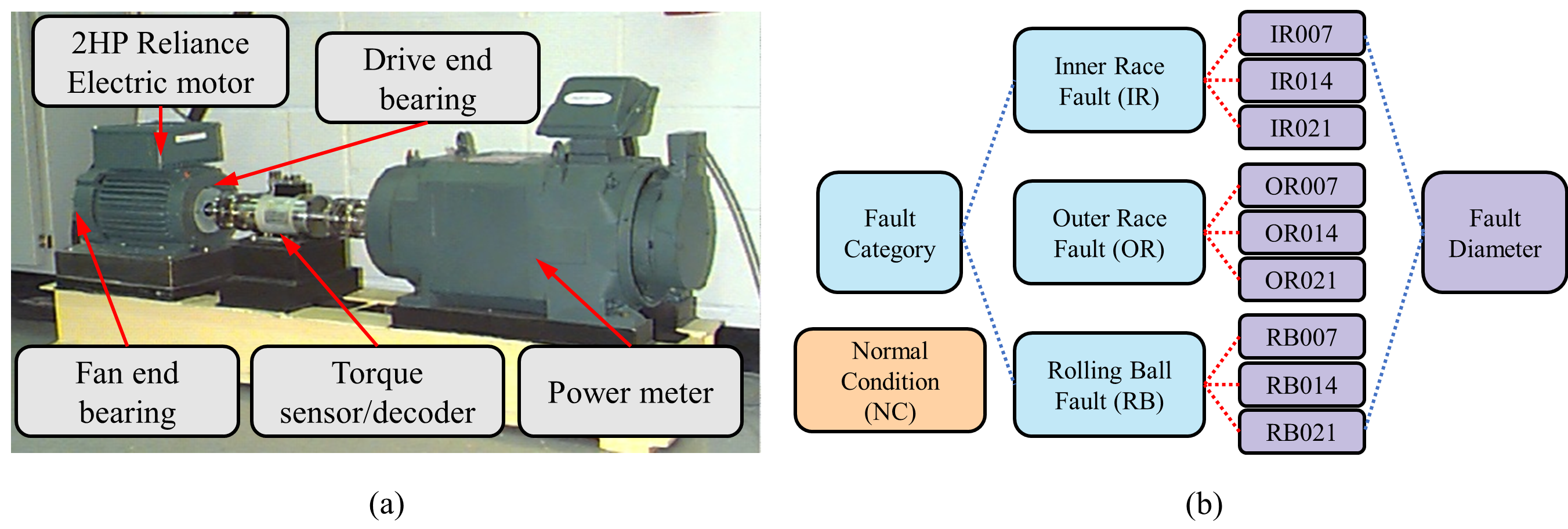}
	  \caption{CWRU dataset introduction (a) experimental device (b) bearing fault modes}
        \label{fig:CWRUdevice}
\end{figure}

In this paper, Case Western Reserve University (CWRU) Bearing Fault dataset is adopted to train and evaluate the proposed TST. The CWRU dataset is a public experimental dataset widely used in the field of rotating machinery fault diagnosis and the experimental device is shown in Fig.\ref{fig:CWRUdevice}(a). The acceleration vibration signals from the fan end bearing (type 6203-2RS JEM SKF) and drive end bearing (type 6205-2RS JEM SKF) are acquired using a 2 horsepower (2HP) reliance electric motor a power meter and torque sensor for different motor load (0HP to 3HP). The motor Speed varies between 1730rpm and 1797rpm. Motor bearings are seeded with faults adopting Electro-Discharge Machining (EDM). There are three bearing fault modes including inner race fault (IR), outer race fault (OR) and rolling ball fault (RB). Besides, different fault diameter ranging from 0.007 inches to 0.021 inches are introduced at IR, OR and RB, separately (shown in Fig.\ref{fig:CWRUdevice}(b)). Healthy bearings without any fault (normal condition, NC) can be regarded as a special fault mode, so there are 10 bearing fault types in the CWRU dataset. Finally, data files are in $.mat$ format and each file contains DE (drive end accelerometer data), FE (fan end accelerometer data), BA (base accelerometer data), time (time series data) and RPM (rpm during testing). 

\begin{figure}[h]
	\centering
		\includegraphics[width = 0.9\textwidth]{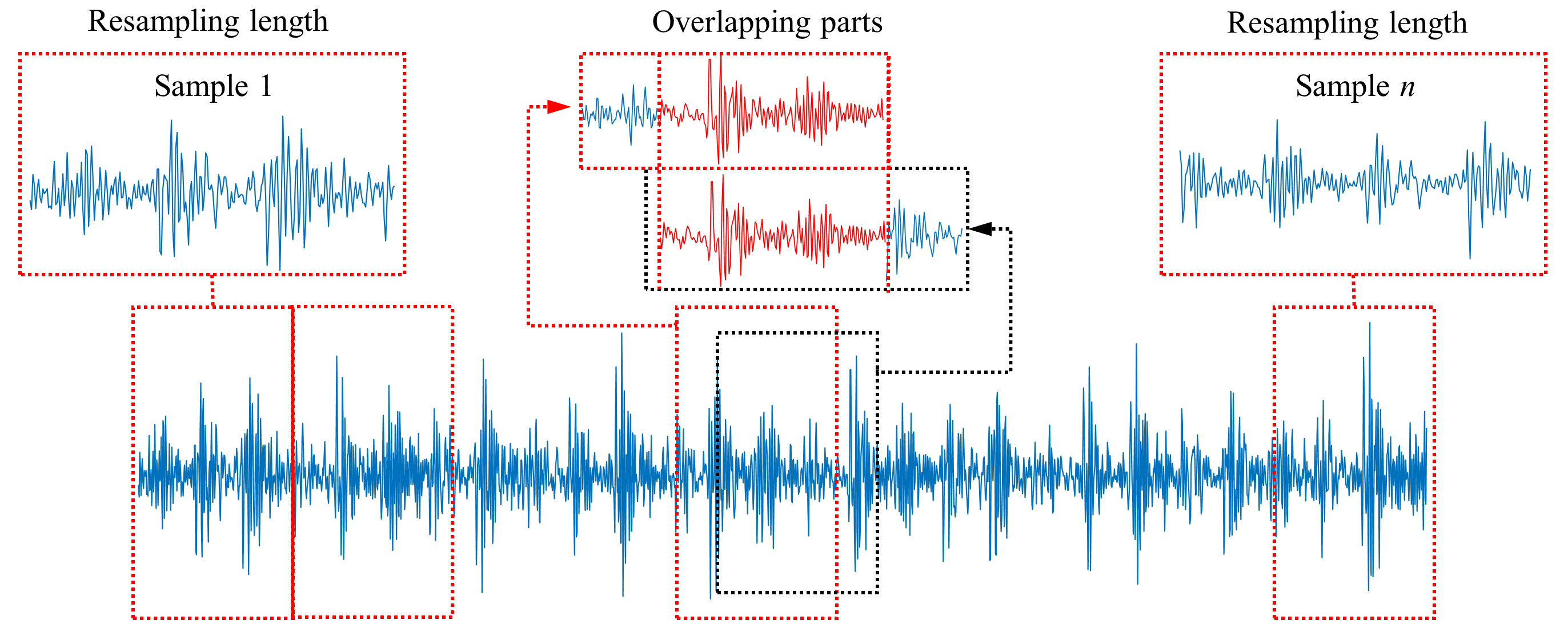}
	  \caption{Resampling process of the CWRU dataset}
        \label{fig:Preprocess}
\end{figure}

\begin{figure}[h]
	\centering
		\includegraphics[width = 0.9\textwidth]{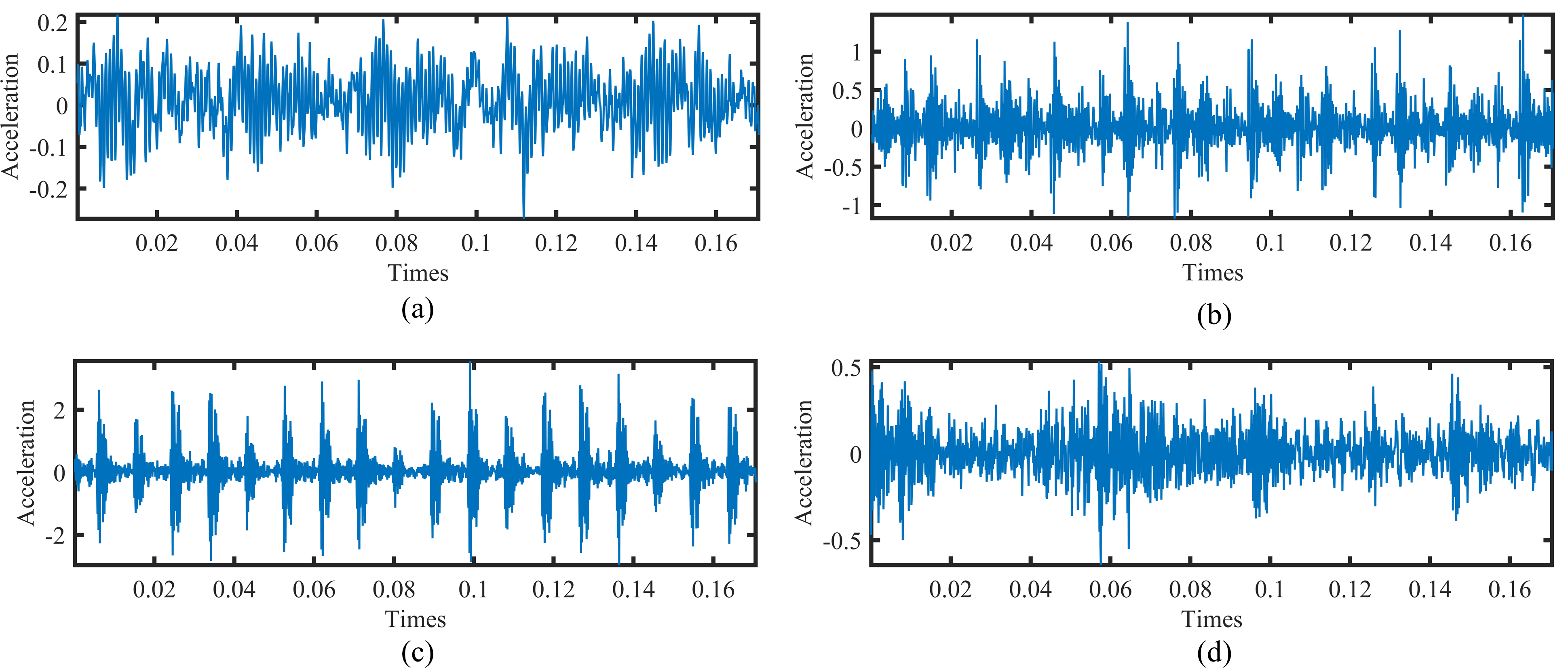}
	  \caption{Part of the dataset via resampling (a) NC (b) IR (c) OR (d) RB}
        \label{fig:SomeDataSet}
\end{figure}

CWRU dataset is collected at 12000 samples/second and 48000 samples/second, which is a very high sampling frequency, so that can not be utilized directly. We employ the resampling method to construct the appropriate dataset in the proposed TST, as shown in Fig.\ref{fig:Preprocess} Raw vibration signals in CWRU dataset are split and trimmed with an applicable resampling length. In this paper, the resampling length is 2048. Besides, a part of the samples are obtained by the delay sampling method for data enhancement. Delay sampling method is to obtain samples with an interval smaller than 2048, which means that there will be some overlapping parts between the samples. This method can produce the function like interpolation. Based on this work, a dataset containing 9000 samples to train and evaluate the proposed TST is obtained. Then, we randomly take 7000 samples as the training set, and the remaining 2000 samples as the test set. Some vibration signals in the dataset are shown in Fig.\ref{fig:SomeDataSet}. It can be found that these signals are messy and irregular, so the fault diagnosis method utilized the TST is needed. 

\section{Experimental results and discussions}
This section will discuss and analyze the fault diagnosis results based on the proposed TST in detail, including the basic fault diagnosis results, comparison with other methods, the influence of the hyperparameters and the visualization of the extracted feature vectors. The hardware environment where the programs run is Intel 10700F, NVIDIA RTX 3070, 32 GB RAM. Besides, Python 3.8, Pytorch 1.8.1 and CUDA 10.2 are adopted for deep learning. 

\subsection{Fault diagnosis results based on the proposed TST and comparison with other methods}
\begin{figure}[h]
	\centering
		\includegraphics[width = 0.99\textwidth]{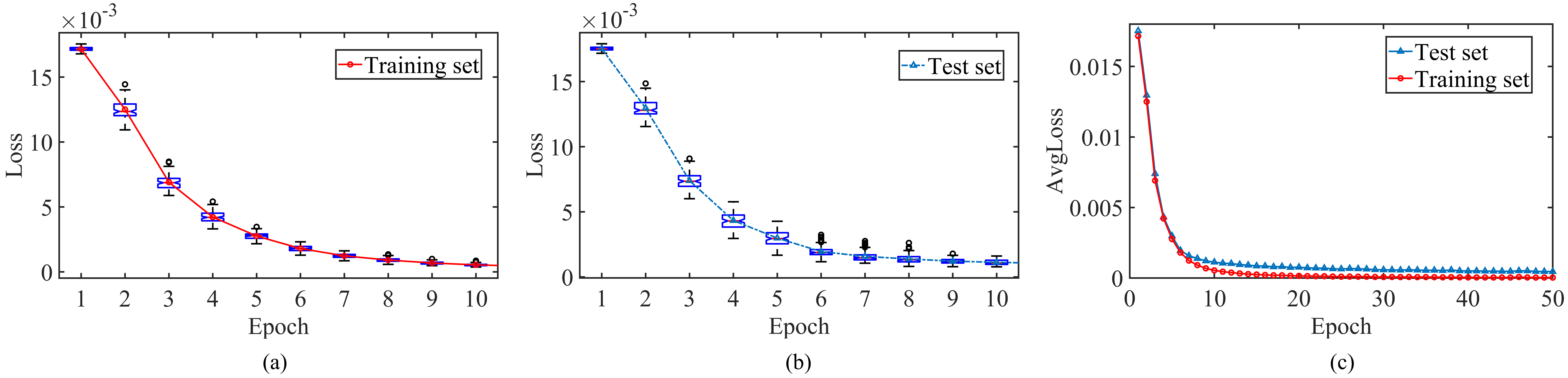}
	  \caption{Loss function value during the training process (a) box plot of the training set in the first 10 epochs (b) box plot of the test set in the first 10 epochs (c) average loss throughout the training process}
        \label{fig:TSTLoss}
\end{figure}

\begin{figure}[h]
	\centering
		\includegraphics[width = 0.99\textwidth]{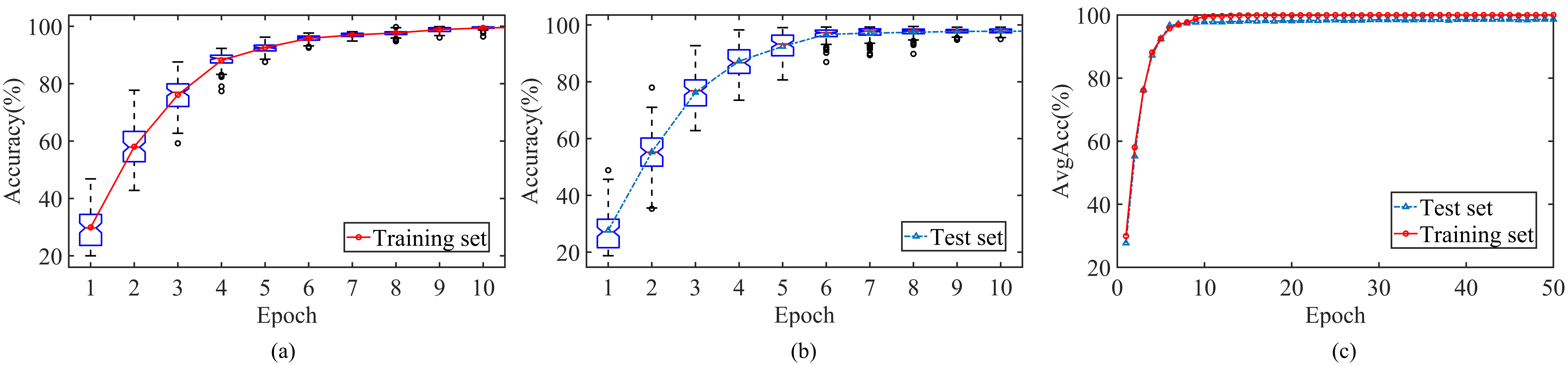}
	  \caption{Accuracy during the training process (a) box plot of the training set in the first 10 epochs (b) box plot of the test set in the first 10 epochs (c) average accuracy throughout the training process}
        \label{fig:TSTAcc}
\end{figure}

During the training process, the batch size is set to 128 and the TST is trained over 50 epochs to extract the robust feature vectors of the bearing fault mode. Moreover, to eliminate the effects of the random initialization, the training process will be repeated 100 times under the same condition. Based on these settings, Fig.\ref{fig:TSTLoss} and Fig.\ref{fig:TSTAcc} show the variation of the loss function and recognition accuracy during the training process, respectively. The box plot with statistical information during the first 10 epochs, the average loss (AvgLoss) and the average recognition accuracy (AvgAcc) in the entire training process are included. 

It can be seen from Fig.\ref{fig:TSTLoss}(a) and Fig.\ref{fig:TSTLoss}(b) that at the early stage of training, the difference of random initialization will lead to a certain range of fluctuations of the loss function values on the training set and the test set. However, after 10 epochs, the box plot become flat, and the deviation range of the outliers (black dot in Fig.\ref{fig:TSTLoss} and Fig.\ref{fig:TSTAcc}) shrinks by degrees, indicating that the training loss and test loss are gradually stabilizing. These results show that the TST can avoid the impact of the random initialization and the performance is robust. Furthermore, the change of AvgLoss throughout the training process is drawn in Fig.\ref{fig:TSTLoss}(c). We can observe that with the process of training, the AvgLoss of the training set and the test set show a gentle downward trend without obvious oscillation. These results illustrate that the optimizer parameters are appropriate and the TST has healthy gradient flow during training. When the training stops, the training AvgLoss and test AvgLoss of the proposed TST are approximately the same, indicating that there is no serious overfitting phenomenon. Moreover, the changing process of the fault diagnosis accuracy is shown in Fig.\ref{fig:TSTAcc}. As can be seen from Fig.\ref{fig:TSTAcc}(a) and Fig.\ref{fig:TSTAcc}(b), similar to the loss function values, the different model initialization parameters at the beginning of training will cause large fluctuations in accuracy. But the training set accuracy and the test set accuracy are basically stable over 10 epochs. Then, Fig.\ref{fig:TSTAcc}(c) shows that the average fault diagnosis accuracy on the training set and test set increase  steadily, which indicates that the TST has good convergence ability under the given parameters. At the end of the 100 times repeated training process, the training set accuracy is 100\% for all the training procedures. In addition, the highest accuracy (TopAcc), the lowest accuracy (MinAcc) and the AvgAcc on the test set of the TST are 99.30\%, 97.25\% and 98.63\%, respectively. These analysis results signify that the proposed TST has great generalization ability and fault diagnosis performance.

\begin{figure}[h]
	\centering
		\includegraphics[width = 0.8\textwidth]{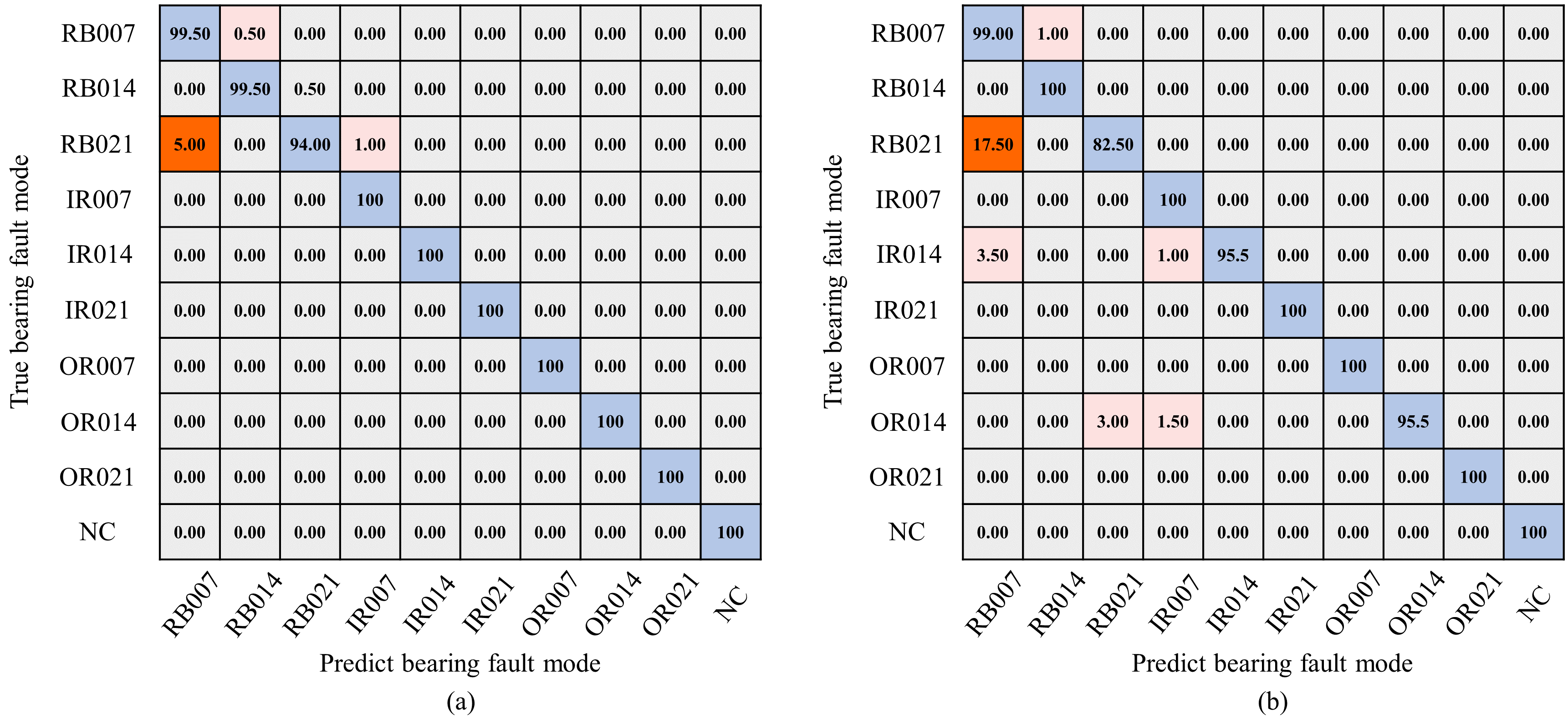}
	  \caption{Confusion matrix of the TST (a) best results (TopAcc=99.30\%) (b) worst results (MinAcc=97.25\%)}
        \label{fig:ConfusionMatrix}
\end{figure}

Then, to further analyze the fault diagnosis results of the TST, Fig.\ref{fig:ConfusionMatrix}(a) and Fig.\ref{fig:ConfusionMatrix}(b) show the confusion matrix of the best and worst fault diagnosis results under the condition of 100 rounds repetitive training process, separately. The rows denote the true fault categories of the samples, and the columns represent the predict bearing fault modes of the TST. In Fig.\ref{fig:ConfusionMatrix}(a), for the best case, the identification accuracy reaches 99.30\% and most of the misclassification categories are concentrated on RB. Besides, the worst results in Fig.\ref{fig:ConfusionMatrix}(b) show a similar pattern, about 17.50\% RB021 samples are misclassified as RB007, indicating that the rolling ball fault is a challenge of the bearing fault diagnosis. However, it can also be noted from Fig.\ref{fig:ConfusionMatrix} that most of the misclassifications are due to the fault diameter, such as RB007 and IR014 are misjudged as RB014 and IR007, rather than the fault mode. This is a natural consequence, because utilizing the given vibration signals to calculate the fault diameter should be a regression problem, not a pattern recognition problem. Moreover, if only the classification accuracy of the four bearing fault modes (IR, OR, RB and NC) is considered, the TopAcc and MinAcc of the TST can achieve 99.90\% and 99.20\%, which is a very brilliant fault diagnosis result. 

Finally, we compare the proposed TST with other ML and DL methods based on the CWRU dataset to further illustrate the effectiveness of the TST. Firstly, according to the adopted dataset format, we design three deep learning models different from the proposed TST for comparison experiments, including: 1) 1D convolutional neural network (CNN1D), which consists of five convolution layers and pooling layers. 2) Recurrent neural network based on the stacked five LSTM layers (RNNLSTM). 3) Recurrent neural network combined with 1D convolution layer (ConvRNN) to better handle 
the long sequence. Of course, the above three models are also carefully tuned for a fair comparison and their details are shown in Tab.\ref{tb:CNNRNN}. Among them, the first convolution layer of the CNN1D utilizes a wide convolution kernel with kernel size=128 to resist noise and extract the global features. In addition, the sequence length of 2048 is too long for the RNN, which will make it difficult for the LSTM unit of the later time step to capture the information from the previous time step. To solve this problem and make the comparative experimental results more convincing, the input of the RNNLSTM in this paper are the 45$\times$45 matrices obtained by reshaping the raw vibration signals (23 points in the samples are ignored). Then, in the ConvRNN, a 1D convolution layer is used to reduce the sequence length of the RNN. Finally, we also compare the fault diagnosis results of our methodology with other researchers' achievements except the models designed by ourselves, including SVM \cite{Deng, Zhang}, BPNN \cite{Jia}, DBN \cite{Shao}, CNN \cite{Levent, Liang, HUANG201977} and GRU \cite{ZhaoRui}. The comparison results can be seen in Tab.\ref{tb:Comparision}. It should be noted that due to the different processing of the dataset, some researchers carry out fault diagnosis tasks with 4 categories (just NC, IR, OR and RB). Therefore, the accuracy of the TST in 4 bearing fault conditions is also given in Tab.\ref{tb:Comparision}. Among these methods, the proposed TST achieves the highest accuracy without relying on additional signal processing algorithms, convincingly demonstrating the superiority of the TST in the bearing fault diagnosis. 
\begin{table}[H]
      \caption{The detailed structure of the comparison models}
      \label{tb:CNNRNN}
      \begin{tabular*}{1.0\textwidth}{@{}LL@{}}
            \toprule
            Models & Structure parameters \\ % Table header row
            \midrule
            CNN1D & $ \left[ {\begin{array}{*{20}{c}}
                  \text{Conv1D(16, 128, 16)} \\
                  \text{BatchNorm} \\
                  \text{ReLU} \\
                  \text{MaxpoolD(2)}
                  \end{array}} \right] \rightarrow \left[ {\begin{array}{*{20}{c}}
                        \text{Conv1D(32, 3, 1)}\\
                        \text{BatchNorm} \\
                        \text{ReLU} \\
                        \text{Maxpool1D(2)}\\
                        \end{array}} \right] \rightarrow \left[ {\begin{array}{*{20}{c}}
                        \text{Conv1D(64, 3, 1)}\\
                        \text{BatchNorm} \\
                        \text{ReLU} \\
                        \text{Maxpool1D(2)} 
                        \end{array}} \right] \times \text{3} \rightarrow \text{Linear(100, ReLU)} $ \\
            RNNLSTM & $ \text{Reshape 45$\times$45} \rightarrow \left[ {\begin{array}{*{20}{c}}
                  \text{LSTM(45, 64, tanh)} \\
                  \text{Dropout(0.1)}
                  \end{array}} \right] \times \text{5} \rightarrow \text{Linear(128, GeLU)} $ \\
            ConvRNN & $ \text{Conv1D(128, 128, 128)} \rightarrow \left[ {\begin{array}{*{20}{c}}
                        \text{LSTM(128, 128, tanh)} \\
                        \text{Dropout(0.1)}
                        \end{array}} \right] \times \text{5} \rightarrow \text{Linear(256, GeLU)} $ \\
            \bottomrule
      \end{tabular*}
\end{table}

\begin{table}[H]
      \caption{Comparison results of the proposed TST and other fault diagnosis methods}\label{tb:Comparision}
      \begin{tabular*}{1.0\textwidth}{@{}LLL@{}}
            \toprule
            Methodology & Signal processing methods & Accuracy \\
            \midrule
            LS-SVM \cite{Deng} & EMD (Empirical Mode Decomposition) and PSO & 89.50\% (4 categories) \\
            ICDSVM \cite{Zhang} & EEMD (Ensemble Empirical Mode Decomposition) & 97.75\% (4 categories) \\
            BPNN \cite{Jia} & Frequency spectra & 81.35\% (10 categories) \\
            DBN \cite{Shao} &  Time-domain features and PSO & 88.20\% (10 categories) \\
            CNN (1D format) \cite{Levent} & Raw vibration signals & 93.30\% (4 categories) \\
            WT-CNN (2D format) \cite{Liang} & Times-frequency domain images by WT (Wavelet Transform) & 95.09\% (4 categories) \\
            MC-CNN (1D format) \cite{HUANG201977} & Raw vibration signals & 98.46\% (4 categories) \\
            RNN \cite{ZhaoRui} & Local features of the time-domain signals & 95.60\% (4 categories) \\
            LFGRU \cite{ZhaoRui} & Local features of the time-domain signals & 99.60\% (4 categories) \\
            CNN1D & Raw vibration signals & 97.32\% (10 categories) \\
            RNNLSTM & Raw vibration signals & 92.07\% (10 categories) \\
            ConvRNN & Raw vibration signals & 94.74\% (10 categories) \\
            \midrule
            TST (Proposed) & Raw vibration signals & \makecell[l]{\textbf{98.63\%} (10 categories) \\ \textbf{99.72\%} (4 categories)} \\
            \bottomrule
      \end{tabular*}
\end{table}

\subsection{Influence of the network structure hyperparameters}
\begin{figure}[h]
	\centering
		\includegraphics[width = 0.99\textwidth]{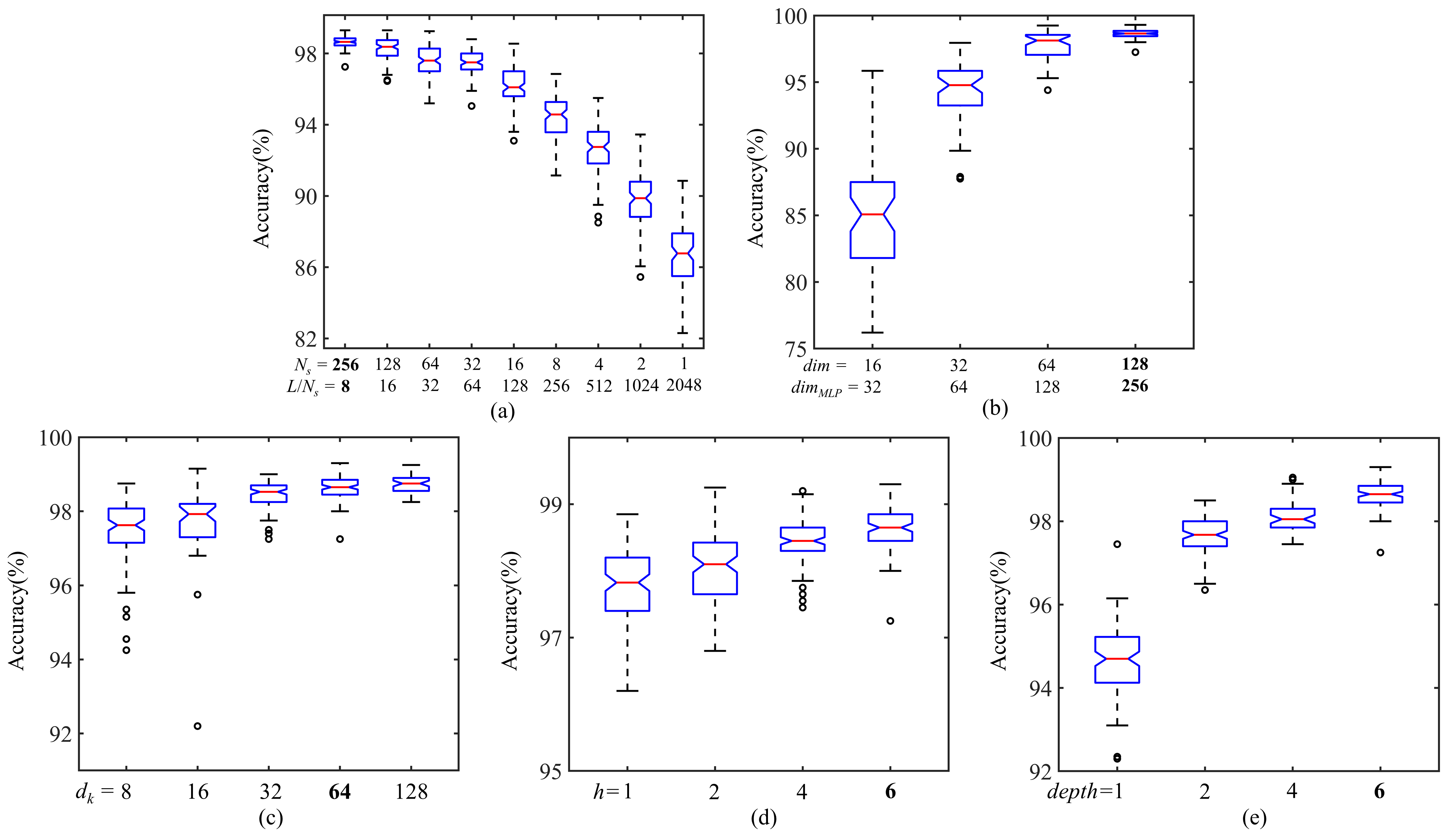}
	  \caption{Box plot of the different hyperparameters (a) $N_{s}$ and $L/(N_{s})$ (b) $dim$ and $dim_{MLP}$ (c) $d_k$ (d) $h$ (e) $depth$}
        \label{fig:BoxplotHyper}
\end{figure}

As described earlier in Section 2, there are many structure hyperparameters such as $N_{s}$, $dim$, $d_{k}$, $h$, $dim_{MLP}$ and $depth$ in the TST. The changes of these hyperparameters may have the impacts on the fault diagnosis performance of the TST, therefore, detailed analysis is required. The hyperparameters are adjusted respectively, then, the fault diagnosis results under the same dataset and training conditions are shown in Tab.\ref{tb:Hyperparameters}. and Fig.\ref{fig:BoxplotHyper}. In addition, the computational complexity (FLOPs) of the TST and the number of learnable parameters under different hyperparameters are also obtained by open source library $thop$ and shown in Tab.\ref{tb:Hyperparameters}. Firstly, it can be seen from Tab.\ref{tb:Hyperparameters} that the subsequence length $L/N_{s}$ can significantly affect the performance of the TST. Compared with CNN, the proposed TST does not have obvious inductive bias such as locality, and translation equivariance. Only MLP is local and translationally equivariant in the TST, while Transformer layer and MSA, as its cores, extract global features. Therefore, too large subsequence length can make the tokens sequence more likely ignore the local features of the vibration signal, which result in the decline of the recognition accuracy. In contrast, smaller $L/N_{s}$ can help the TST better retain the local features of the original input by the time series embedding and MLP, meanwhile, MSA can extract and express the global features, so that the TST has stronger feature extraction capability. The results in Fig.\ref{fig:BoxplotHyper}(a) confirm this point, and simultaneously, we can also observe that the subsequence length affects the stability of the TST. In the baseline model, TopAcc=99.30\%, MinAcc=97.25\%, AvgAcc=98.63\% and standard deviation (Std) is 0.31\%, indicating that random initialization has no significant effect on the model performance. Whereas, when $L/N_{s}$ is 2048, TopAcc, MinAcc, AvgAcc and Std are 90.85\%, 82.30\%, 86.73\% and 1.76\%, respectively. Besides, the statistical box plot in Fig.\ref{fig:BoxplotHyper}(a) becomes more slender, which illustrates that the long subsequence length will make the model unstable. Then, an interesting analysis result can also be obtained from Tab.\ref{tb:Hyperparameters} that the computation complexity of the TST can be greatly reduced by using a smaller $N_{s}$, but its parameters number has not changed significantly. For example, changing $N_{s}$ from 256 to 1 reduces the computational complexity by nearly 99.14\% (405.52MFLOPs to 3.41MFLOPs), however, the parameters number increases by only 16.46\% (1.58M to 1.84M). This is because the part of the TST mainly consuming the computing resources is MSA, and it can be seen from Eq.\eqref{eq:SDPAttn} and Eq.\eqref{eq:MHA} that the computational complexity of MSA is directly related to $(N_{s}+1)$, but its calculation process does not contain any learnable parameters. The difference in the number of model parameters under various $N_{s}$ is only reflected in the procedures of the time series embedding and position encoding, which is a small value compared to the entire model size. 
\begin{table}[h]
      \caption{Influence of the TST structure hyperparameters}\label{tb:Hyperparameters}
      \begin{tabular*}{1.0\textwidth}{@{}LLLLLLLLLLLL@{}}
            \toprule
            \multirow{2}*{Label} & \multicolumn{8}{c}{Hyperparameters} & \multirow{2}*{AvgAcc} & \multirow{2}*{FLOPs} & \multirow{2}*{\makecell[l]{Parameters \\ number}} \\
            ~ & $N_{s}$ & $L/N_{s}$ & $dim$ & $dim_{MLP}$ & $d_{k}$ & $h$ & $depth$ & Pos encoding & ~ \\
            \midrule
            Baseline & 256 & 8 & 128 & 256 & 64 & 6 & 6 & 1D & \textbf{98.63\%} & 405.52M & 1.58M \\
            \midrule
            \multirow{8}*{A} & 128 & 16 & ~ & ~ & ~ & ~ & ~ & ~ & 98.25\% & 203.18M & 1.58M \\
            ~ & 64 & 32 & ~ & ~ & ~ & ~ & ~ & ~ & 97.58\% & 102.51M & 1.58M \\
            ~ & 32 & 64 & ~ & ~ & ~ & ~ & ~ & ~ & 97.53\% & 52.17M & 1.59M \\
            ~ & 16 & 128 & ~ & ~ & ~ & ~ & ~ & ~ & 96.20\% & 27.00M & 1.59M \\
            ~ & 8 & 256 & ~ & ~ & ~ & ~ & ~ & ~ & 94.38\% & 14.42M & 1.61M \\
            ~ & 4 & 512 & ~ & ~ & ~ & ~ & ~ & ~ & 92.73\% & 8.13M & 1.64M \\
            ~ & 2 & 1024 & ~ & ~ & ~ & ~ & ~ & ~ & 89.73\% & 4.98M & 1.71M \\
            ~ & 1 & 2048 & ~ & ~ & ~ & ~ & ~ & ~ & 86.73\% & 3.41M & 1.84M \\
            \midrule
            \multirow{3}*{B} & ~ & ~ & 16 & 32 & ~ & ~ & ~ & ~ & 85.16\% & 39.51M & 0.15M \\
            ~ & ~ & ~ & 32 & 64 & ~ & ~ & ~ & ~ & 94.36\% & 82.18M & 0.32M \\
            ~ & ~ & ~ & 64 & 128 & ~ & ~ & ~ & ~ & 97.76\% & 177.00M & 0.69M \\
            \midrule
            \multirow{3}*{C} & ~ & ~ & ~ & ~ & 8 & ~ & ~ & ~ & 97.50\% & 139.25M & 0.55M \\
            ~ & ~ & ~ & ~ & ~ & 16 & ~ & ~ & ~ & 97.67\% & 177.15M & 0.69M \\
            ~ & ~ & ~ & ~ & ~ & 32 & ~ & ~ & ~ & 98.46\% & 252.94M & 0.99M \\
            ~ & ~ & ~ & ~ & ~ & 128 & ~ & ~ & ~ & \textbf{98.74\%} & 707.69M & 2.76M \\
            \midrule
            \multirow{3}*{D} & ~ & ~ & ~ & ~ & ~ & 1 & ~ & ~ & 97.67\% & 151.88M & 0.60M \\
            ~ & ~ & ~ & ~ & ~ & ~ & 2 & ~ & ~ & 98.03\% & 202.41M & 0.79M \\
            ~ & ~ & ~ & ~ & ~ & ~ & 4 & ~ & ~ & 98.46\% & 303.47M & 1.19M \\
            \midrule
            \multirow{3}*{E} & ~ & ~ & ~ & ~ & ~ & ~ & 1 & ~ & 94.65\% & 67.67M & 0.27M \\
            ~ & ~ & ~ & ~ & ~ & ~ & ~ & 2 & ~ & 97.67\% & 135.04M & 0.53M \\
            ~ & ~ & ~ & ~ & ~ & ~ & ~ & 4 & ~ & 98.11\% & 269.78M & 1.05M \\
            \midrule
            F & ~ & ~ & ~ & ~ & ~ & ~ & ~ & None & 96.11\% & 404.52M & 1.55M \\
            \bottomrule
      \end{tabular*}
\end{table}

Furthermore, as can be seen from the Tab.\ref{tb:Hyperparameters} and Fig.\ref{fig:BoxplotHyper}(b), $dim$ and $dim_{MLP}$ are also key factors in determining the TST recognition accuracy. It should be noted that $dim$ and $dim_{MLP}$ need to be adjusted together and in this paper, $dim_{MLP}=2dim$ is set. Similar to the word embedding procedure in the NLP, time series embedding is to treat each subsequence as a "word" and map it to a higher-dimensional embedding space by linear transformation. Then, the linearly separable high dimensional features are obtained throughout the Transformer layer. However, small $dim$ and $dim_{MLP}$ will limit the dimension of the embedding space, which may make it more difficult for the TST to learn the linearly separable features. In terms of the model stability, embedding dimension shows a pattern similar to the subsequence length. As shown in Fig.\ref{fig:BoxplotHyper}(b), a smaller $dim$ means a thinner box plot, which indicates that too small embedding dimension will reduce the stability of the model. In the aspect of the model parameters and computational complexity, it can be seen from Eq.\eqref{eq:MHA} and Eq.\eqref{eq:MLP} that the dimensions of $W_{embedding}$, $W_{1}^{l}$ and $W_{2}^{l}$ are related to $dim$ and $dim_{MLP}$. Therefore, the FLOPs and parameters number of the TST are distinctly affected by them. Afterwards, from the Tab.\ref{tb:Hyperparameters}, Fig.\ref{fig:BoxplotHyper}(c) and Fig.\ref{fig:BoxplotHyper}(d), one unanticipated finding is that the dimension of keys $d_{k}$ and the number of $Head$ in Eq.\eqref{eq:MHA} have no serious impress on the accuracy and stability of the model. A possible explanation for this might be that the substance of Scaled Dot-Product Attention in Eq.\eqref{eq:SDPAttn} is to calculate the similarity between the tokens sequence based on the vector dot product, and linear transformations using different $d_{k}$ may not change it significantly. Moreover, MSA can abstract the global features in different embedding dimensions compared with the general attention mechanism, but this advantage may not be obvious in the given dataset. In addition, we can notice that when $d_{k}$=128, the accuracy of the TST is 0.11\% higher than the baseline model. Nevertheless, due to the huge computational complexity (707.69MFLOPs), we do not take it as the baseline model. After that, the number of the Transformer basic blocks $depth$ is also an important hyperparameter. From Fig.\ref{fig:BoxplotHyper}(e) we can see that a deeper Transformer layer within limits predicates a more stable training process and higher accuracy, which is an apparent result. Meanwhile, this result can also prove that the residual connection can prevent the TST from degradation, like ResNet \cite{ResNet}. Of course, greater $depth$ is not always better. Too large $depth$ will enormously increase the computational complexity and parameters number of the model, and may lead to the over-fitting problem.

Finally, the necessity of the position coding is also confirmed in Tab.\ref{tb:Hyperparameters}. The accuracy of the TST will decrease by approximately 2.52\% without the position coding. This is because the position information of the subsequences is not retained during the entire calculation process of the TST, so that the model needs to learn a "jigsaw puzzle" procedure by itself and the absence of the position encoding will make this learning process more difficult. These results are in line with those of previous studies in reference \cite{ViT}. 

\subsection{Visualization of the proposed TST}
In addition to the final recognition accuracy, the distribution form of the feature vectors extracted by the model in the embedding space is also a meaningful indicator to evaluate the generalization ability of the model. At the same time, observing the variation of feature vectors can also help us to better explain the work pattern of the model. Therefore, in this section, we will focus on discussing the visualization of the proposed TST. 

\begin{figure}[h]
	\centering
		\includegraphics[width = 0.99\textwidth]{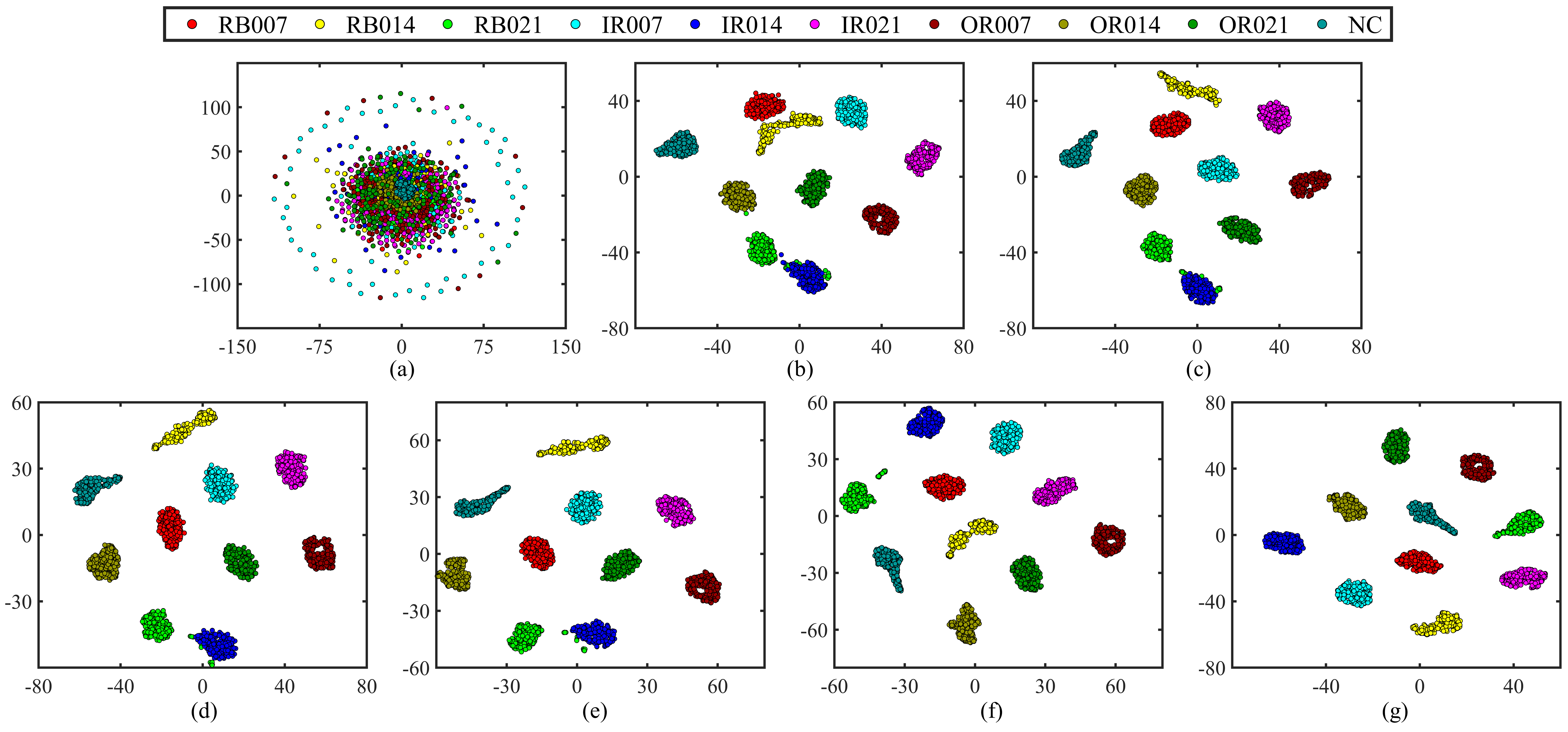}
	  \caption{Visualization of the class tokens in different Transformer basic block via t-SNE (a) raw vibration signals (b)-(g) 1st-6th Transformer basic block}
        \label{fig:TransformerVisual}
\end{figure}

Firstly, to analyze the distribution of the feature vectors, visualization of the embedding space in the TST is necessary. However, as shown in previous contents, the embedding dimension of the baseline model is 128, which is a higher-dimensional space that cannot be visualized directly. Consequently, dimensionality reduction methods are required. The dimensionality reduction methods that are widely adopted in practical problems include Principal Component Analysis (PCA), Multiple Dimensional Scaling (MDS) and t-distributed Stochastic Neighbor Embedding (t-SNE). Among them, PCA and MDS are linear dimension reduction methods, which may fail for complex high-dimensional data. t-SNE, based on manifold learning, is one of the best dimension reduction and visualization methods. This paper also utilizes t-SNE to reduce the dimensionality of feature vectors. The feature map generation procedure of TST is similar to BERT, and the feature vectors are the class tokens in Eq.\eqref{eq:ClassToken}. Fig.\ref{fig:TransformerVisual} provides the visualization results of the raw vibration signals and class tokens in each Transformer basic block via t-SNE. From Fig.\ref{fig:TransformerVisual}(a) we can see a large overlapping area in the visualization result of the original signals, and there is no obvious boundaries between the different fault categories. Hence, it is hard to straightly make classification based on the raw vibration signals. Then, the baseline model in this paper is stacked with six Transformer basic blocks, however, surprisingly, the various categories of class tokens in the first layer have shown separability to a degree. This finding illustrates the powerful feature extraction capability of the TST. But the results are not brilliant enough because there is some confusion between RB021 (green points) and IR014 (blue points). Besides, the decision boundary between RB014 (yellow points) and RB007 (red points) is too close, which can easily lead to the domain shift problem. So a deeper network architecture is needed. The results of the correlational analysis are shown in Fig.\ref{fig:TransformerVisual}(c)-(f). In the 2nd Transformer basic block, the confusion degree in class tokens of RB021 and IR014 is reduced, and their decision boundaries becomes clearer. In the meantime, the inter-class separability among RB014 and other categories also becomes better. A semblable pattern is found in Fig.\ref{fig:TransformerVisual}(d) and (e). Nevertheless, there are still a few number of RB021 samples clearly falling within the decision boundary of IR014, until 5th Transformer basic block. It can be seen from the data in Fig.\ref{fig:TransformerVisual}(f) that there is no significant overlapping part among various categories of samples, but some dispersion of the class tokens in the same fault mode can be observed, such as RB021. Finally, the distribution of the class tokens in 6th Transformer basic block (that is, the feature vectors of the TST) is drawn in Fig.\ref{fig:TransformerVisual}(g). We can see great intra-class compactness and inter-class separability of the feature vectors in disparate fault modes from Fig.\ref{fig:TransformerVisual}(g), indicating the effectiveness of the proposed TST. 

\begin{figure}[h]
	\centering
		\includegraphics[width = 0.99\textwidth]{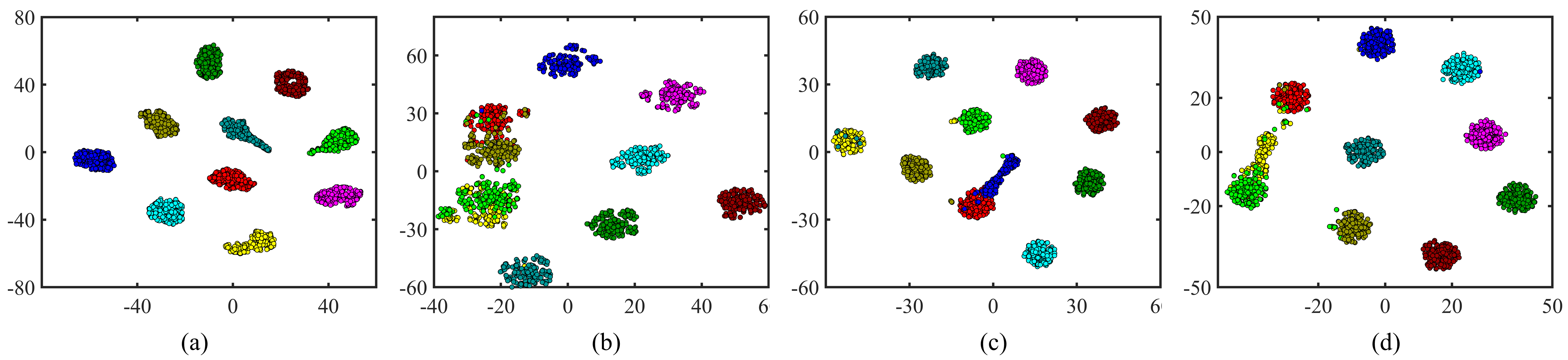}
	  \caption{Visualization of the feature vectors extracted by different models (a) TST (AvgAcc=98.63\%) (b) RNNLSTM (AvgAcc=92.07\%) (c) ConvRNN (AvgAcc=94.74\%) (d) CNN1D (AvgAcc=97.32\%)}
        \label{fig:VisualCompare}
\end{figure}

Moreover, Fig.\ref{fig:VisualCompare} compares the distribution form of feature vectors extracted by different models. The detailed structural parameters of these models are shown in Tab.\ref{tb:Comparision}. The analysis results illustrate that among these deep learning models, the feature vectors obtained by the TST possess the best intra-class compactness and inter-class separability, so that the TST achieves prime fault diagnosis accuracy and  generalization ability (as shown in Fig.\ref{fig:VisualCompare}(a)). On the contrary, the visualization results in Fig.\ref{fig:VisualCompare}(b) of the RNNLSTM present that there is a distinct overlapping region between the feature vectors of some different categories, such as RB014 and RB021, which is the reason for the low fault identification accuracy. Furthermore, the feature vectors of the same category in RNNLSTM are also relatively dispersed, that is, the intra-class compactness is not valid enough, which will lead to the poor generalization ability of the model. Finally, it can be seen from the data in Fig.\ref{fig:VisualCompare}(c) and Fig.\ref{fig:VisualCompare}(d) that the feature vectors extracted by ConvRNN and CNN1D show good inter-class separability, but the decision boundaries of some fault modes are too close, including RB007 and IR014 in Fig.\ref{fig:VisualCompare}(c), RB014 and RB021 in Fig.\ref{fig:VisualCompare}(d). Unclear decision boundaries also can reduce the generalization ability of the model, therefore, the fault diagnosis performance of ConvRNN and CNN1D is inferior to that of TST. These findings further prove that TST has better feature extraction capability compared with traditional deep learning models based on the CNN and RNN architecture.

\section{Conclusions}
In this paper, a new fault diagnosis model which is able to process 1D format data directly based on the Time Series Transformer has been established, and the effectiveness of the proposed model has been verified utilizing the CWRU dataset. Then, the influence of some structure hyperparameters on the model performance, computational complexity and parameters number has been analyzed in detail. Finally, the distribution form of feature vectors has been presented. The main conclusions are as follows.

(1) The proposed TST can deal with the raw vibration signals directly without additional signal processing algorithm. The feature extraction process of TST only employs the MSA, and completely abandons the basic operations in CNN or RNN architecture. The experimental results on the CWRU dataset show that the fault recognition accuracy of TST reaches 98.63\%, and its performance surpasses many traditional deep learning models such as RNNLSTM, ConvRNN and CNN1D. 

(2) Hyperparameters such as the number of subsequences ($N_{s}$), time series embedding dimension ($dim$), the dimension of MLP ($dim_{MLP}$) and queries ($d_{k}$), the number of head ($h$), the number of Transformer basic blocks ($depth$) and position encoding will affect the performance of TST. Among them, smaller subsequence length, larger $dim$ and $dim_{MLP}$ can improve the recognition accuracy of TST and enhance the stability of the training process, but it also means greater computational complexity. Then, $d_{k}$ has no significant impact on the performance and stability of the model because the linear transformation in different dimensions may not seriously change the similarity between two tokens sequences. The effect of $h$ shows a similar pattern, and a possible explanation for this might be that the advantage of extracting features from different dimensions may not be obvious under the given dataset. Besides, due to the existence of residual connections, the proposed TST avoids degradation when $h$ increases. Finally, position coding is necessary. 

(3) t-SNE method is adopted to visualize the feature vectors in the embedding space. It is found that the feature vectors extracted by TST have better intra-class compactness and inter-class separability compared with RNNLSTM, ConvRNN and CNN1D. The effectiveness of the proposed method is further proved. 

In the future work, further research should focus on the following aspects. Firstly, it is too rough to obtain the tokens sequence just by once linear transformation, and multiple convolution operations can make this process more elaborate and adjustable. Some researches in CV field have proved that introducing convolution operations in Transformer can enhance the performance of the model. Secondly, the TST proposed in this paper is based on the vanilla Transformer architecture. At present, deep learning researchers have proposed many more advanced Transformer architectures such as PVT, T2TViT. These modified models can also be transferred to the fault diagnosis field. 

\section*{Acknowledgements}
It is very grateful for the financial supports from the National Major Science and Technology Projects of China (No. 2017-IV-0008-0045) and the National Natural Science Foundation of China (Nos. 11972129, 11732005).

% \label{}

% Use the style of numbering in square brackets.
% If nothing is used, default style will be taken.
%\begin{enumerate}[a)]
%\item 
%\item 
%\item 
%\end{enumerate}  

%\begin{itemize}
%\item 
%\item 
%\item 
%\end{itemize}  

%\begin{description}
%\item[]
%\item[] 
%\item[] 
%\end{description}  

% Figure
% \begin{figure}[<options>]
% 	\centering
% 		\includegraphics[<options>]{}
% 	  \caption{}\label{fig1}
% \end{figure}

% \begin{table}[<options>]
% \begin{tabular*}{\tblwidth}{@{}LL@{}}
% \toprule
% \midrule
%  & \\
%  & \\
%  & \\
%  & \\
% \bottomrule
% \end{tabular*}
% \end{table}

% Uncomment and use as the case may be
%\begin{theorem} 
%\end{theorem}

% Uncomment and use as the case may be
%\begin{lemma} 
%\end{lemma}


% \section{}
% \label{}

\printcredits

\bibliographystyle{model1-num-names}
% \bibliographystyle{cas-model2-names}

\bibliography{cas-refs}



\end{document}